\documentclass[usenatbib]{mn2e}
\usepackage{apjfonts}
\usepackage{graphicx}

\newcommand{\araa}{ARA\&A}   
\newcommand{\aj}{AJ}         
\newcommand{\aap}{A\&A}      
\newcommand{\aaps}{A\&AS}    
\newcommand{\aapr}{A\&ARv}   
\newcommand{\apj}{ApJ}       
\newcommand{\apjs}{ApJS}     
\newcommand{\apjl}{ApJ}      
\newcommand{\mnras}{MNRAS}   
\newcommand{\nat}{Nature}    
 
\def\picplace#1{\vbox{\hrule\@height 0.4pt\@width\hsize
\hbox to\hsize{\vrule\@width 0.4pt\@height#1\hfil
\vrule\@width 0.4pt\@height#1}\hrule\@height 0.4pt\@width\hsize}}
\def\squareforqed{\hbox{\rlap{$\sqcap$}$\sqcup$}}
\def\sq{\ifmmode\squareforqed\else{\unskip\nobreak\hfil
\penalty50\hskip1em\null\nobreak\hfil\squareforqed
\parfillskip=0pt\finalhyphendemerits=0\endgraf}\fi}

\def\la{\mathrel{\mathchoice {\vcenter{\offinterlineskip\halign{\hfil
$\displaystyle##$\hfil\cr<\cr\sim\cr}}}
{\vcenter{\offinterlineskip\halign{\hfil$\textstyle##$\hfil\cr
<\cr\sim\cr}}}
{\vcenter{\offinterlineskip\halign{\hfil$\scriptstyle##$\hfil\cr
<\cr\sim\cr}}}
{\vcenter{\offinterlineskip\halign{\hfil$\scriptscriptstyle##$\hfil\cr
<\cr\sim\cr}}}}}
\def\ga{\mathrel{\mathchoice {\vcenter{\offinterlineskip\halign{\hfil
$\displaystyle##$\hfil\cr>\cr\sim\cr}}}
{\vcenter{\offinterlineskip\halign{\hfil$\textstyle##$\hfil\cr
>\cr\sim\cr}}}
{\vcenter{\offinterlineskip\halign{\hfil$\scriptstyle##$\hfil\cr
>\cr\sim\cr}}}
{\vcenter{\offinterlineskip\halign{\hfil$\scriptscriptstyle##$\hfil\cr
>\cr\sim\cr}}}}}

\def\utw{\smash{\rlap{\lower5pt\hbox{$\sim$}}}}
\def\udtw{\smash{\rlap{\lower6pt\hbox{$\approx$}}}}

\def\diameter{{\ifmmode\mathchoice
{\ooalign{\hfil\hbox{$\displaystyle/$}\hfil\crcr
{\hbox{$\displaystyle\mathchar"20D$}}}}
{\ooalign{\hfil\hbox{$\textstyle/$}\hfil\crcr
{\hbox{$\textstyle\mathchar"20D$}}}}
{\ooalign{\hfil\hbox{$\scriptstyle/$}\hfil\crcr
{\hbox{$\scriptstyle\mathchar"20D$}}}}
{\ooalign{\hfil\hbox{$\scriptscriptstyle/$}\hfil\crcr
{\hbox{$\scriptscriptstyle\mathchar"20D$}}}}
\else{\ooalign{\hfil/\hfil\crcr\mathhexbox20D}}%
\fi}}

\def\bbbc{{\mathchoice {\setbox0=\hbox{$\displaystyle\rm C$}\hbox{\hbox
to0pt{\kern0.4\wd0\vrule height0.9\ht0\hss}\box0}}
{\setbox0=\hbox{$\textstyle\rm C$}\hbox{\hbox
to0pt{\kern0.4\wd0\vrule height0.9\ht0\hss}\box0}}
{\setbox0=\hbox{$\scriptstyle\rm C$}\hbox{\hbox
to0pt{\kern0.4\wd0\vrule height0.9\ht0\hss}\box0}}
{\setbox0=\hbox{$\scriptscriptstyle\rm C$}\hbox{\hbox
to0pt{\kern0.4\wd0\vrule height0.9\ht0\hss}\box0}}}}
\def\bbbq{{\mathchoice {\setbox0=\hbox{$\displaystyle\rm
Q$}\hbox{\raise
0.15\ht0\hbox to0pt{\kern0.4\wd0\vrule height0.8\ht0\hss}\box0}}
{\setbox0=\hbox{$\textstyle\rm Q$}\hbox{\raise
0.15\ht0\hbox to0pt{\kern0.4\wd0\vrule height0.8\ht0\hss}\box0}}
{\setbox0=\hbox{$\scriptstyle\rm Q$}\hbox{\raise
0.15\ht0\hbox to0pt{\kern0.4\wd0\vrule height0.7\ht0\hss}\box0}}
{\setbox0=\hbox{$\scriptscriptstyle\rm Q$}\hbox{\raise
0.15\ht0\hbox to0pt{\kern0.4\wd0\vrule height0.7\ht0\hss}\box0}}}}
\def\bbbt{{\mathchoice {\setbox0=\hbox{$\displaystyle\rm
T$}\hbox{\hbox to0pt{\kern0.3\wd0\vrule height0.9\ht0\hss}\box0}}
{\setbox0=\hbox{$\textstyle\rm T$}\hbox{\hbox
to0pt{\kern0.3\wd0\vrule height0.9\ht0\hss}\box0}}
{\setbox0=\hbox{$\scriptstyle\rm T$}\hbox{\hbox
to0pt{\kern0.3\wd0\vrule height0.9\ht0\hss}\box0}}
{\setbox0=\hbox{$\scriptscriptstyle\rm T$}\hbox{\hbox
to0pt{\kern0.3\wd0\vrule height0.9\ht0\hss}\box0}}}}
\def\bbbs{{\mathchoice
{\setbox0=\hbox{$\displaystyle     \rm S$}\hbox{\raise0.5\ht0\hbox
to0pt{\kern0.35\wd0\vrule height0.45\ht0\hss}\hbox
to0pt{\kern0.55\wd0\vrule height0.5\ht0\hss}\box0}}
{\setbox0=\hbox{$\textstyle        \rm S$}\hbox{\raise0.5\ht0\hbox
to0pt{\kern0.35\wd0\vrule height0.45\ht0\hss}\hbox
to0pt{\kern0.55\wd0\vrule height0.5\ht0\hss}\box0}}
{\setbox0=\hbox{$\scriptstyle      \rm S$}\hbox{\raise0.5\ht0\hbox
to0pt{\kern0.35\wd0\vrule height0.45\ht0\hss}\raise0.05\ht0\hbox
to0pt{\kern0.5\wd0\vrule height0.45\ht0\hss}\box0}}
{\setbox0=\hbox{$\scriptscriptstyle\rm S$}\hbox{\raise0.5\ht0\hbox
to0pt{\kern0.4\wd0\vrule height0.45\ht0\hss}\raise0.05\ht0\hbox
to0pt{\kern0.55\wd0\vrule height0.45\ht0\hss}\box0}}}}
\def\bbbz{{\mathchoice {\hbox{$\sf\textstyle Z\kern-0.4em Z$}}
{\hbox{$\sf\textstyle Z\kern-0.4em Z$}}
{\hbox{$\sf\scriptstyle Z\kern-0.3em Z$}}
{\hbox{$\sf\scriptscriptstyle Z\kern-0.2em Z$}}}}

\newcommand{\HST}{\emph{HST}}
\newcommand{\Modot}{M_{\odot}}

\newcommand{\gI}{{\rm F475W}\!-\!{\rm F814W}}

\newcommand{\cM}{{\cal{M}}}
\newcommand{\lamc}{\Lambda_{\rm c}}

\newcommand{\mybibitem}[3]{\bibitem[\protect\citeauthoryear{#1}{#2}]{#3}}
\newcommand{\mybibthree}[4]{\bibitem[\protect\citeauthoryear{#1}{#2}{#3}]{#4}}
\begin{document}

\title[SGB and RC morphologies in eMSTO clusters]{On the
  interpretation of sub-giant branch morphologies of intermediate-age star
  clusters with extended main sequence turnoffs}   

\author[P.\ Goudfrooij et al.]{Paul Goudfrooij,$^{1}$\thanks{E-mail: goudfroo@stsci.edu}
L\'eo Girardi,$^{2}$ Philip Rosenfield,$^{3}$ Alessandro Bressan,$^{4}$ 
\newauthor 
Paola Marigo,$^{3}$ Matteo Correnti,$^{1}$ 
and Thomas H.\ Puzia$^5$ \smallskip  
\\ 
$^1$\,Space Telescope Science Institute, 3700 San Martin Drive,
 Baltimore, MD 21218, U.S.A. \\
$^2$\,Osservatorio Astronomico di Padova -- INAF, Vicolo dell'Osservatorio 5,
  I-35122 Padova, Italy\\
$^3$\,Dipartimento di Fisica e Astronomia Galileo Galilei, Universit\`a di
  Padova, Vicolo dell'Osservatorio 3, I-35122 Padova, Italy\\
$^4$\,SISSA, via Bonomea 365, I-34136 Trieste, Italy \\
$^5$\,Institute of Astrophysics, Pontificia Universidad Cat\'{o}lica de Chile,
  Av.\ Vicu\~{n}a Mackenna 4860, 7820436 Macul, Santiago, Chile
}

\date{Accepted 2015 March 25. Received 2015 March 10; in original form
  2015 February 5}  

\maketitle

\begin{abstract}
Recent high-quality photometry of many star clusters in the Magellanic
Clouds with ages of 1\,--\,2 Gyr revealed main sequence turnoffs (MSTOs) that
are significantly wider than can be accounted for by a simple stellar population
(SSP). 
Such extended MSTOs (eMSTOs) are often interpreted in terms of an age
spread of several 10$^8$ yr, challenging the traditional view of star
clusters as being formed in a single star formation episode. 
Li et al. and Bastian \& Niederhofer recently investigated the sub-giant
branches (SGBs) of NGC~1651, NGC~1806, and NGC~1846, 
three star clusters in the Large Magellanic Cloud (LMC) that exhibit an eMSTO. 
They argued that the SGB of these star clusters can be explained only by a SSP.  
We study these and two other similar star clusters in the LMC, using extensive
simulations of SSPs including unresolved binaries.  We find that the
shapes of the cross-SGB profiles of all star 
clusters in our sample are in fact consistent with their cross-MSTO profiles
when the latter are interpreted as age distributions. Conversely, SGB
morphologies of star clusters with eMSTOs are found to be inconsistent
with those of simulated SSPs. 
Finally, we create PARSEC isochrones from tracks featuring a grid
  of convective overshoot levels and a very fine grid of stellar
  masses. A comparison of the observed photometry with these
  isochrones shows that the morphology of the red clump (RC) of such
  star clusters is also consistent with that implied by their MSTO in
  the age spread scenario.  
We conclude that the SGB and RC morphologies of star clusters featuring
eMSTOs are consistent with the scenario in which the eMSTOs are caused by a
distribution of stellar ages. 
\end{abstract}



\section{Introduction}
\label{s:intro}

Until recently, globular clusters (GCs) were thought to be simple stellar
populations (hereafter SSPs), consisting of thousands to millions of coeval
stars with the same chemical composition. In the past decade, however, a 
consensus has emerged that GCs typically harbor multiple stellar populations
featuring several unexpected characteristics 
\citep[e.g.,][and references therein]{bedi+04,piot+07,grat+12}. 
Multiple sequences in several major features of 
colour-magnitude diagrams (CMDs) are now commonplace in massive Galactic star
clusters, e.g, in their main sequence (MS), sub-giant branch (SGB), and red
giant branch (RGB). These multiple sequences are especially evident in
\emph{Hubble Space Telescope (HST)} photometry using colours
involving passbands that cover molecular features of OH, NH, and CN in the
near-ultraviolet part of the electromagnetic spectrum \citep[see,
e.g.,][]{piot+12,bell+13,milo+12a,milo+13,milo+15,dott+15}. 

Meanwhile, recent spectroscopic surveys established that light elements such as N, O,
and Na show large star-to-star abundance variations (often dubbed
``Na-O anticorrelations'') within virtually all Galactic GCs studied
to date in sufficient detail \citep[][and references therein]{carr+10}.  
The chemical processes involved in causing the light-element abundance
variations have largely been identified as proton capture reactions at high
temperature ($T \ga 2\times 10^7$ K), such as the CNO and NeNa cycles. 
Currently, the leading candidates for polluting sources (``polluters'') are
stars in which such reactions occur readily and which feature slow stellar winds
so that their ejecta are relatively easy to retain within the potential well of
massive clusters: (1) intermediate-mass asymptotic giant branch (AGB) and
super-AGB stars ($4 \la {\cal{M}}/M_{\odot} \la 10$; e.g., \citealt{danven07}),
(2) rapidly rotating massive stars (often referred to as ``FRMS'';
\citealt{decr+07}) and (3) massive binary stars \citep{demi+09}.  

In the two currently favored formation scenarios, these chemical
anticorrelations are due to stars having either formed from or polluted by gas
that is a mixture of  pristine material and material shed by such polluters.
In the ``extended  
star formation'' scenario \citep[see, e.g.,][]{derc+08,conspe11,valcat11}, the
abundance variations are caused by a second generation of stars that formed out
of gas clouds that were polluted by winds of first-generation stars to varying
extents, during a period spanning up to a few hundreds of Myr, depending on
the nature of the polluters and the depth of the potential well. The
alternative ``early disc accretion'' scenario 
\citep{bast+13a} does \emph{not} involve extended star formation: the polluted
gas is instead produced by FRMS and massive binary stars, and accreted by
low-mass pre-main-sequence stars during the first $\approx$\,20 Myr after the
formation of the star cluster.  

In the context of the nature of Na-O anticorrelations in Galactic GCs, the
recent discovery of extended main sequence turn-offs (hereafter eMSTOs) in
inter\-me\-diate-age (1\,--\,2 Gyr old) star clusters in the Magellanic Clouds
\citep*{macbro07,mack+08a,glat+08,milo+09,goud+09,goud+11a,goud+11b} has
generated much interest. Several investigations suggested 
that the eMSTOs are due to the presence of multiple stellar populations spanning
an age interval of several $10^8$ yr within these clusters \citep*[see
also][]{rube+10,rube+11,kell+11,kell+12,gira+13a,corr+14,goud+14}.     
The leading alternative theory for the cause of the eMSTO phenomenon is that it
is due to spreads in rotation velocity among turnoff stars (hereafter the ``stellar
rotation'' scenario:\ \citealt*{basdem09,li+12,yang+13,li+14}, but see
\citealt*{gira+11}). 

Due to the important implications of the answer to the question whether or not 
eMSTOs in intermediate-age clusters are due to a significant range in stellar
ages, the issue is being studied from a variety of angles. Recent
investigations looked for evidence of extended star formation (or a
significant lack thereof) in young massive star clusters in the Large
Magellanic Cloud (LMC), with seemingly conflicting results (see
\citealt{bassil13}  
versus \citealt{corr+15}). Others studied relations between properties of
eMSTOs and dynamical properties of the clusters, finding a correlation between
eMSTO width and cluster escape velocity which seems to be understood most
straightforwardly in terms of the extended star formation scenario 
(\citealt{goud+14}, hereafter G+14). 
However, results with the opposite conclusion were recently reported by
\citeauthor{li+14} (2014, hereafter L+14) and \citeauthor{basnie15} (2015,
hereafter BN15). L+14 studied the SGB morphology of
NGC~1651, an eMSTO cluster of age $\simeq$\,2 Gyr, and concluded that it can
be explained only by a single-age stellar population. Showing results of new
calculations of stellar tracks in the MSTO-SGB region of the CMD with vs.\ 
without stellar rotation, L+14 argued that eMSTOs in intermediate-age clusters
are most likely caused by a range of stellar rotation velocities within such
clusters. BN15 performed a similar study of the SGB morphologies in NGC~1806 and
NGC~1846, two other eMSTO clusters, again concluding that age spreads are
unlikely to be the cause of the eMSTO phenomenon. 

In the current paper, we investigate the claims of L+14 and BN15 by means of an
independent study of the SGB properties of five intermediate-age star clusters
in the LMC (including the clusters studied by L+14 and BN15). 
After a brief description of the data and SSP models used in this
paper in Sect.\ \ref{s:data}, we discuss properties of cross-SGB
magnitude distributions for these star clusters and highlight differences with
the methodologies used by L+14 and BN15 in Sect.\ \ref{s:SGBmorph}. Sect.\ \ref{s:disc}
reviews the relevance of convective overshoot to the morphology of the CMD of
intermediate-age clusters, and discusses our cross-SGB distributions when
compared with a set of isochrones in which the implementation of the dependence
of convective overshoot on stellar mass is different from that used in the
isochrones used by L+14 and BN15. 
Sect.\ \ref{s:disc} also includes an investigation of the claim by BN15
  that the morphologies of the red clumps in these star clusters are
  inconsistent with the age spread scenario. 
Our conclusions and their implications are summarized in Sect.\ \ref{s:conc}. 

\section{Data and Models}
\label{s:data}

We use the \HST\ photometry of intermediate-age star clusters in the LMC that
was described in detail in \citet[][hereafter G+11a]{goud+11a}, G+14, and
\citet{corr+14}. Briefly, the data from G+11a used here involves
multi-exposure photometry taken with the Wide Field Camera (WFC) of the
Advanced Camera for Surveys (ACS) using the F435W and F814W filters (\HST\
program 10595, PI: P. Goudfrooij). The data from G+14 used here involves
similar photometry, now taken with the Wide Field Camera \#3 (WFC3) using the
F475W and F814W filters (\HST\ program 12257, PI: L. Girardi), while the data
from \citet{corr+14} used here involves ACS/WFC photometry using the F555W and
F814W filters with one exposure per filter (\HST\ ``snapshot'' program 9891,
PI: G. Gilmore).   

\begin{figure*}
\centerline{\includegraphics[width=17cm]{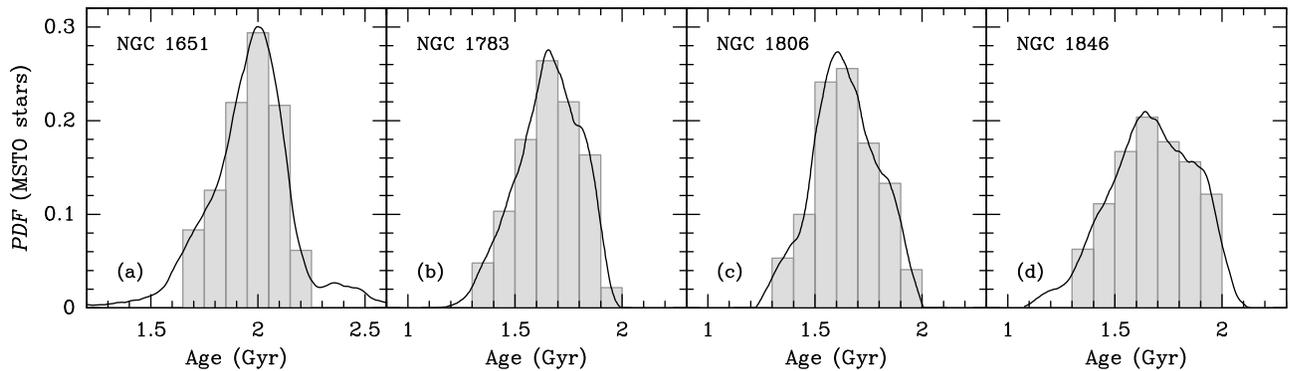}}
\caption{Illustration of the assignment of relative masses to individual SSP
  ages (i.e., isochrones) for the ``multi-SSP'' simulations of the 4 star
  clusters in our sample that feature eMSTOs (see Sect.\ \ref{s:meth}). 
  \emph{Panel (a)}: The black solid line represents the pseudo-age distribution
  of NGC 1651 taken from G+14, expressed as probability density function
  (PDF). The histogram in light grey shows the relative weights of the different
  age bins used during the simulations.  
  \emph{Panel (b)}: Same as Panel (a), but now for NGC~1783, using its 
  pseudo-age distribution taken from G+11a, using ages from the M+08
  isochrones. 
  \emph{Panel (c)}: Same as Panel (b), but now for NGC~1806.  
  \emph{Panel (d)}: Same as Panel (b), but now for NGC~1846, using its 
  pseudo-age distribution taken from \citet{goud+09}. 
  } 
\label{f:weights}
\end{figure*}

From this sample we first select the star clusters in the LMC for which the
eMSTO feature is clearly established. Specifically, we select star clusters
for which G+14 measured full-width at half maximum (FWHM) values of their MSTO
cross-cut of at least 350 Myr when expressed as an age range (i.e., their
``pseudo-age distributions''). Since we are interested in measuring the SGB
morphologies of such clusters, we subselect clusters for which their CMD
includes at least 35 stars in the SGB region\footnote{We found that clusters
  with fewer than 30\,--\,35 SGB stars showed significant stochastic
  fluctuations in their cross-SGB magnitude distributions (see Sect.\
  \ref{s:SGBmorph}).}. 
This selection procedure yields the three massive clusters NGC 1783, NGC 1806,
and NGC 1846. To this sample we add NGC 1651 (which also features an eMSTO), the
cluster for which \citet{li+14} claimed that its SGB morphology can be explained
only by a single-age stellar population. Finally, we add IC 2146, a cluster
whose mass and age are very similar to those of NGC 1651 although it does
\emph{not} show evidence for an eMSTO. We find that the SGB properties of IC
2146 reveal information that is relevant to the question whether or not (and, if
so, how) SGB morphologies of eMSTO clusters can be used to rule out extended
star formation periods. Relevant properties of the star clusters in our sample
are listed in Table~\ref{t:sample}. 

\begin{table}
\footnotesize
\begin{center}
\caption[]{Properties of star clusters in our sample.}
\label{t:sample}
\begin{tabular}{@{}r@{~~~~}ccc@{~~~}c@{~~}c@{~~}c@{}} \hline \hline
\multicolumn{3}{c}{~~} \\ [-2.5ex]  
\multicolumn{1}{c}{Name} & log $\cM_{\rm cl}$ & $r_c$ & $r_{\rm eff}$ & Ref. & Age & Ref. \\ [0.2ex]
\multicolumn{1}{c}{(1)} & (2) & (3) & (4) & (5) & (6) & (7) \\ [0.3ex] \hline 
\multicolumn{3}{c}{~~} \\ [-1.5ex] 
NGC 1651 & $4.91 \pm 0.06$ & $ 4.57 \pm 0.36$ & \llap{1}$2.82 \pm 2.01$ & 
 1 & 2.00 & 1 \\ 
NGC 1783 & $5.42 \pm 0.11$ & \llap{1}$0.50 \pm 0.49$ & \llap{1}$1.40 \pm 2.24$ &
 1 & 1.70 & 2 \\
NGC 1806 & $5.10 \pm 0.06$ & $ 5.91 \pm 0.27$ & $ 9.04 \pm 1.24$ & 
 1 & 1.65 & 2 \\
NGC 1846 & $5.24 \pm 0.09$ & $ 8.02 \pm 0.49$ & $ 8.82 \pm 0.68$ & 
 1 & 1.65 & 3 \\ 
 IC 2146 & $4.49 \pm 0.07$ & $ 8.89 \pm 1.36$ & \llap{1}$2.53 \pm 1.92$ & 
 4 & 1.90 & 4 \\ [0.5ex] \hline
\multicolumn{5}{c}{~~} \\ [-1.8ex]
\end{tabular}                   
\parbox{8.3cm}{{\bf Notes.}~~Columns: (1) Name of star cluster. (2) Logarithm of
  cluster mass in $\Modot$. (3) Core radius in pc. (4) Effective radius in
  pc. (5) Reference for cluster mass and radii (1 = G+14, 2 = G+11a, 
  3 = \citealt{goud+09}, 4 = \citealt{corr+14}). (6) 
  Age in Gyr used for the ``single-SSP'' simulations in Sect.\
  \ref{s:SGBmorph}. (7) Reference for age data. 
}
\end{center}
\end{table}

To avoid issues related to the presence of a significant number of LMC field
stars in relevant parts of the CMDs, we only consider stars within the core
radius of the clusters. The only exception to this is the case of NGC~1651
for which the combination of cluster mass and core radius would yield too few
stars in the SGB region to produce statistically robust results. Hence we
use all stars within the effective radius for NGC~1651. The influence of field
stars in the SGB region is negligible in all cases, as can be appreciated from
the various CMD plots (panels (a) in Figs.\ \ref{f:n1651cmds} and
\ref{f:SGB2146}\,--\,\ref{f:SGB1846}). 

We adopt the isochrones of \citet[][hereafter M+08]{mari+08} for much of our
analysis in this paper, since these isochrones were also used by G+11a, G+14,
L+14, \citet{corr+14}, and BN15. For each cluster, we adopt the values for $Z$
(always 0.008), distance $(m-M)_0$, best-fitting age, and foreground reddening
($A_V$) found by G+11a, G+14, and \citet{corr+14}.  
The population properties of these clusters were determined
  from the brightnesses and colors of the MSTO and the RGB bump, and the slope
  of the RGB. The SGB morphology was not involved in the determination of
  cluster ages.

\section{SGB Morphologies}
\label{s:SGBmorph}

\subsection{Method}
\label{s:meth}

To test whether the SGB morphologies of the clusters in this paper are
best described by a single-age SSP or by a distribution of ages similar to those
indicated by their MSTO morphologies (i.e., the pseudo-age distributions shown
in G+11a and G+14), we create cross-SGB magnitude distributions (hereafter
called ``cross-SGB profiles'') of the observations and compare them with such
profiles derived from a set of Monte Carlo simulations of SSPs (including
unresolved binary stars) as described below. 

\begin{figure*}
\centerline{\includegraphics[width=13cm]{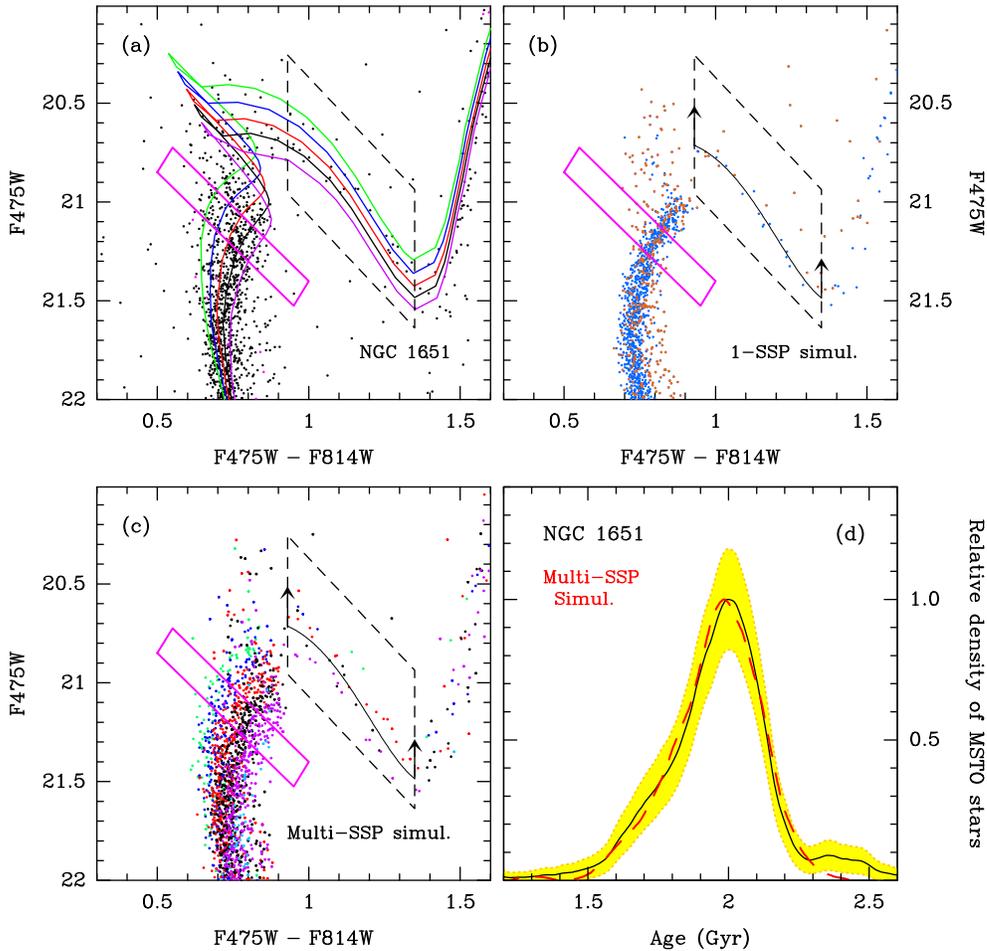}}
\caption{\emph{Panel (a)}: the CMD of NGC 1651 in the MSTO-SGB region (taken
  from G+14) along with \citet{mari+08} isochrones for $Z = 0.008$ and
  ages 1.7, 1.8, 1.9, 2.0, and 2.1 Gyr. 
  Black dots represent stars within the effective radius of NGC~1651. 
  Red dots indicate stars in the outskirts of the \HST\ image with a total
  area equal to that within the effective radius of NGC~1651.  
  The SGB selection box is outlined by black dashed lines. For
  reference, the MSTO crosscut parallelogram used by G+14 to measure the
  pseudo-age distribution of NGC 1651 is shown in magenta. 
  \emph{Panel (b)}: CMD of a single-SSP simulation of NGC 1651 using the 2.0
  Gyr isochrone (see Sect.\ \ref{s:meth}). Single stars are shown as blue
  dots while unresolved binaries are shown as orange dots. The black solid
  line within the SGB selection box in panel (b) represents the
  SGB portion of the isochrone for the best-fitting (mean) age.
  The black arrows in panel (b) indicate the direction of the positive X axis in
  the cross-SGB profiles shown in Figures~\ref{f:SGB1651}\,--\,\ref{f:SGB1846}.    
  \emph{Panel (c)}: similar to panel (b) except that the dots now 
  represent a ``multi-SSP'' simulation of NGC~1651, described in Sect.\
  \ref{s:meth}. 
  Simulated stars with ages of 1.7, 1.8, 1.9, 2.0, 2.1, and 2.2 Gyr are
    shown in green, blue, red, black, purple, and cyan, respectively. 
  \emph{Panel (d)}: the black solid line
  represents the pseudo-age distribution of NGC 1651 as measured from
  the MSTO crosscut by G+14 (cf.\ Fig 1a), expressed as a normalised
  density function, while the yellow region indicates its 68\% confidence
  interval. For comparison, the red dashed line represents the
  corresponding pseudo-age distribution of the ``multi-SSP''
  simulation described in Sect.\ \ref{s:meth} for NGC~1651. 
  } 
\label{f:n1651cmds}
\end{figure*}

Candidate SGB stars are selected by means of a parallelogram in the CMD,
illustrated in panels (a)\,--\,(c) of Fig.~\ref{f:n1651cmds} by black
dashed lines. The blue edge of the parallelogram is placed at a colour
that is clearly redder than the MSTO feature, while the red edge of
the parallelogram is placed at the colour reached by the SGB stars
with the minimum luminosity in the relevant isochrones. The placement
of the top and bottom of the parallelogram is guided by the
distribution of stars in the various SSP simulations, making sure all
artificial SGB stars are captured (including unresolved binaries) and
verifying that the cluster's SGB stars are captured as well.  
Cross-SGB profiles are then created by measuring every star's
magnitude offset from the isochrone with the best-fitting (mean) age as
established in G+11a, G+14, and \citet{corr+14} at the colour of the star in
question, using linear interpolation in colour space.

SSP simulations are conducted by populating isochrones with stars randomly
drawn from a \citet{salp55} IMF between the minimum and maximum stellar masses
in the isochrone.  
We add unresolved binary companions to a fraction of the stars 
using the binary fractions found by G+11a, G+14, and \citet{corr+14}, in
conjunction with a flat primary-to-secondary mass ratio distribution. Finally,
we add random photometric errors whose dependence on the location in the CMD
was derived from the artificial star tests described by G+11a, G+14, and
\citet{corr+14}.   

We perform two distinct sets of SSP simulations for each cluster: (1) a set of 
``single SSP'' simulations, using the isochrone with the best-fitting (mean) age; 
(2) a set of ``multi-SSP'' (or ``composite SSP'') simulations. These
use the isochrones whose ages encompass the range of ages indicated by
the clusters' pseudo-age distributions that were derived from MSTO
crosscut profiles by G+11a and G+14. We employ an age resolution of 0.1
Gyr. For each age, the relative number of stars used in the simulation is set
by the median amplitude of the pseudo-age distribution for that age bin, in
probability density units (this procedure is illustrated 
in Figure~\ref{f:weights})\footnote{This procedure is somewhat different from
  that used by BN15, who fitted single gaussians to the clusters' binned MSTO
  crosscut profiles.}.  
The level to which such ``multi-SSP'' simulations approximate the
relevant CMD features of the star cluster in question is illustrated in
Fig.~\ref{f:n1651cmds} for the case of NGC~1651. In particular, panel (d) of
Fig.~\ref{f:n1651cmds} shows a comparison of the pseudo-age
distribution of NGC~1651 as derived from its MSTO crosscut profile by G+14
(i.e., the curve shown also in Fig.~\ref{f:weights}a) with that of its
``multi-SSP'' set of simulations. 
The overall total number of stars in both sets of SSP simulations is normalised
to the number of cluster stars on the CMD brighter than the 50\% completeness 
limit. 
Finally, all individual simulations are repeated 20 times and their cross-SGB
profiles are averaged together.

\subsection{Results}
\label{s:results}

\subsubsection{NGC 1651}

\begin{figure}
\centerline{\includegraphics[width=7.0cm]{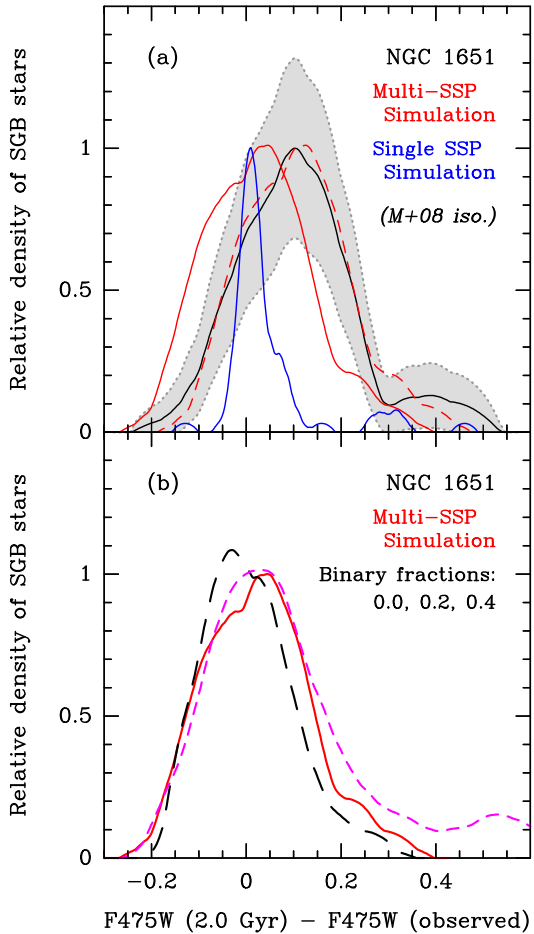}}
\caption{\emph{Panel (a)}: The black solid line represents the cross-SGB profile  
  for NGC 1651, while the light grey region indicates its 68\% confidence
  interval. The red solid line represents the cross-SGB profile of the
  ``multi-SSP'' simulation described in Sect.\ \ref{s:results}.1. The red
  dashed line is the same after shifting it by 0.08 mag to the right. For
  comparison, the solid blue line represents the cross-SGB profile of a
  single-SSP simulation of NGC 1651, divided by a factor of $\simeq
  3.5$ in order to scale its maximum density to that of NGC~1651. 
  \emph{Panel (b)}: Comparison between ``multi-SSP'' simulations of NGC
    1651 created with binary fractions 0.0 (long dashed line), 0.2 (solid line),
    and 0.4 (short dashed line).
  } 
\label{f:SGB1651}
\end{figure}

The resulting cross-SGB profiles for NGC 1651 are shown in
Figure~\ref{f:SGB1651}a. The profiles were derived as density
functions, using the non-parametric Epanechnikov kernel \citep{silv86}. This
minimizes biases that can arise if fixed bin widths are used. 

The cross-SGB profile for the cluster itself (black solid line) shows some
significant differences with that of the single-SSP simulation (blue
line): the cross-SGB profile of the cluster is significantly wider than that of
the single-SSP simulation, and there is an offset between the two profiles in
that the cluster profile is centered on a F475W magnitude that is
$\simeq$\,0.08 mag brighter than that of the single-SSP
simulation. The same offset is present between the profiles of the cluster and
that of its ``multi-SSP'' simulation (see red solid line), while the \emph{width}
and shape of the latter are consistent with that of the cluster's
profile (see dashed red line, which is offset by 0.08 mag from the solid red
line in the positive X direction). \emph{The similarity of the shapes of the
cross-SGB profiles of the cluster and that of its multi-SSP simulation, along
with the profile of the single-SSP simulation being significantly narrower,
strongly suggests that the SGB morphology of NGC 1651 is better described by a
distribution of ages similar to that implied by its cross-MSTO profile, rather
than by a single SSP}. This conclusion is opposite 
to that of L+14, who stated that the cross-SGB profile of NGC 1651
``can be explained only by a single-age stellar population, even though the
cluster has a clearly extended main-sequence turn-off region'' (quoting their 
abstract). Since both conclusions were drawn from the same
original dataset, we discuss this disagreement in some detail below. 

The analysis methods of L+14 differ in two main ways from those
of ours. One difference is that L+14 did not use SSP simulations
in their analysis. The ages in their cross-SGB profiles (their Fig.\ 4) were
determined directly from the location of the isochrones in the CMD,
thus only involving single stars. Conversely, we use SSP simulations
including measurement errors and unresolved binary stars in this
context. We note that the binary fractions of intermediate-age star
clusters like NGC 1651 are substantial ($\approx$\,20\%,
\citealt{mack+08a,milo+09}; G+11a; G+14), and our simulations show
that binaries extend the luminosity distribution of the SGB towards
brighter magnitudes  
at the ages of such clusters (see, e.g., Fig.\ \ref{f:n1651cmds}b). 
The effect of the binary fraction on the cross-SGB profile shape is
  illustrated in Fig.~\ref{f:SGB1651}b for binary fractions of 0\%, 20\%, and
  40\%. 
Secondly and perhaps more importantly, L+14 did not take
the stars in the SGB region of the CMD located above their log\,[age\,(yr)] =
9.24 isochrone into account (see their Figs.\ 2 and 4). 
Given our results shown in Fig.\ \ref{f:SGB1651}a, we suspect that if they
would have done so,  
they would have found that the cross-SGB profile of NGC 1651 continues up
to brighter magnitudes (equivalent to younger ages) with a profile shape 
that is clearly wider than that of a single SSP. 

Finally, the \emph{offset} in magnitude between the cross-SGB profile of NGC 1651 and
that of its SSP simulations is also likely relevant to the disagreement between the
conclusions of L+14 and ours, since L+14 based their conclusions in part on
the observation that the age distribution measured from the cross-SGB profile
of NGC 1651 reached an age lower than that measured from its cross-MSTO
profile (cf.\ their Fig.\ 4). This will be discussed below. 

\begin{figure}
\centerline{\includegraphics[width=7.0cm]{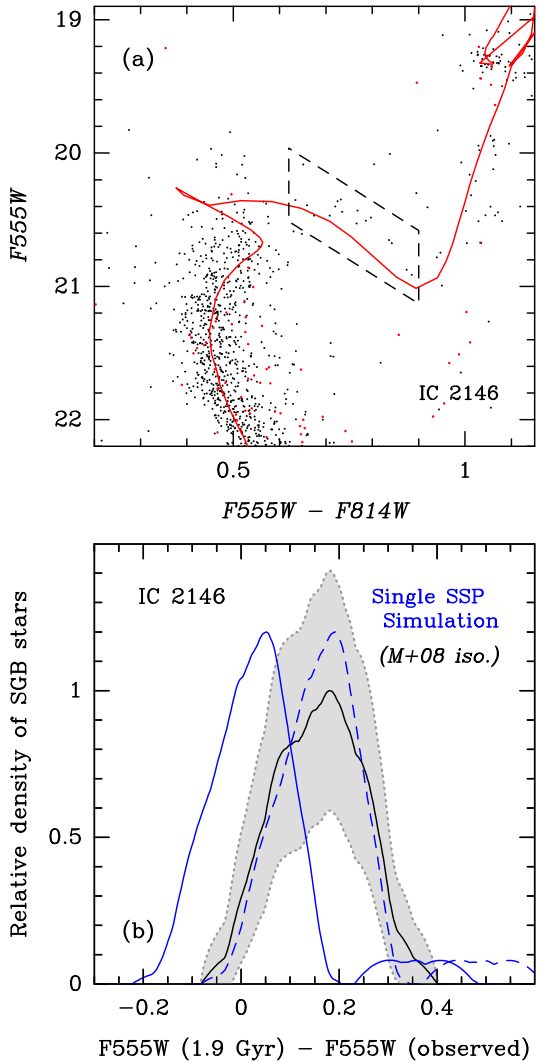}}
\caption{\emph{Panel (a)}: CMD of IC~2146 within its core radius,
  taken from \citet{corr+14}, along with a M+08 isochrone of age 1.90
  Gyr. The SGB selection box is outlined by black dashed lines. 
  Red dots indicate stars in the outskirts of the \HST\ image with a total
  area equal to that within the core radius of IC~2146.  
  \emph{Panel (b)}: The black solid line represents the cross-SGB profile 
  for IC 2146, while the light grey region indicates its 68\% confidence
  interval. For comparison, the solid blue line represents the cross-SGB
  profile of a single-SSP simulation of IC 2146. The blue 
  dashed line is the same after shifting it by 0.13 mag to the right. 
  } 
\label{f:SGB2146}
\end{figure}

\subsubsection{IC 2146}

We recall that this cluster was included in our sample because its mass and
age are similar to those of NGC 1651; while NGC~1651 features an
eMSTO, IC~2146 does not\footnote{G+14 suggested that this difference is due to
  their different escape velocities.}. As such it provides a relevant
comparison in the context of the nature of the SGB morphologies of such clusters. 

Figure~\ref{f:SGB2146}b shows a comparison between the cross-SGB profiles of
IC~2146 (black line) and that of its single-SSP simulation (blue solid
line). Note that in this case, the width of the profile of the single-SSP
simulation is very similar to that of the cluster itself. This difference with
respect to NGC 1651 is consistent with IC~2146 not hosting an eMSTO
whereas NGC~1651 does, although this may in principle also be due in part to
the larger  photometric errors in the dataset of IC~2146 relative to those of
the other clusters presented in this paper. However, an important result of
the simulations of IC~2146 is that \emph{the magnitude offset between the
  cross-SGB profiles of NGC 1651 and its SSP simulation (cf.\ above) is also
  present for IC~2146}. This suggests that the cause of this offset is
\emph{not} related to the question whether or not an intermediate-age cluster
hosts an eMSTO (or a distribution of stellar ages). Instead, we suggest that the offset
is of a systematic nature, related to one or more parameters relevant to
the MSTO-SGB era of stellar evolution theory for stars of moderate metallicity and
masses of $\approx$\,1.5 M$_{\odot}$. This is explored further in Sect.\
\ref{s:disc}.   

\begin{figure}
\centerline{\includegraphics[width=7.1cm]{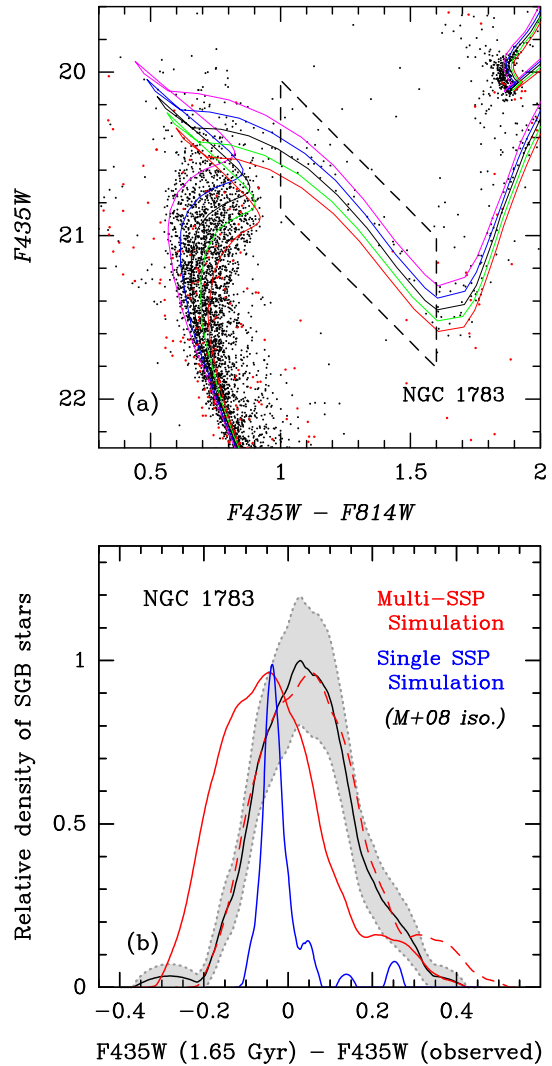}}
\caption{\emph{Panel (a)}: CMD of NGC 1783 within its core radius,
  taken from G+11a, along with M+08 isochrones of age 1.50, 1.60,
  1.70, 1.80, and 1.90 Gyr (top to bottom). 
  Red dots indicate stars in the outskirts of the \HST\ image with a total
  area equal to that within the core radius of NGC~1783.  
  The SGB selection box is
  outlined by black dashed lines. \emph{Panel (b)}:  
  Similar to Figure~\ref{f:SGB1651}, but now for NGC 1783. The SGB
  profile of the ``Single SSP Simulation'' was divided by a factor of $\simeq
  4.5$ in order to scale its maximum density to that of NGC~1783.
  } 
\label{f:SGB1783}
\end{figure}

\subsubsection{NGC 1783, NGC 1806, and NGC 1846}

\begin{figure}
\centerline{\includegraphics[width=7.0cm]{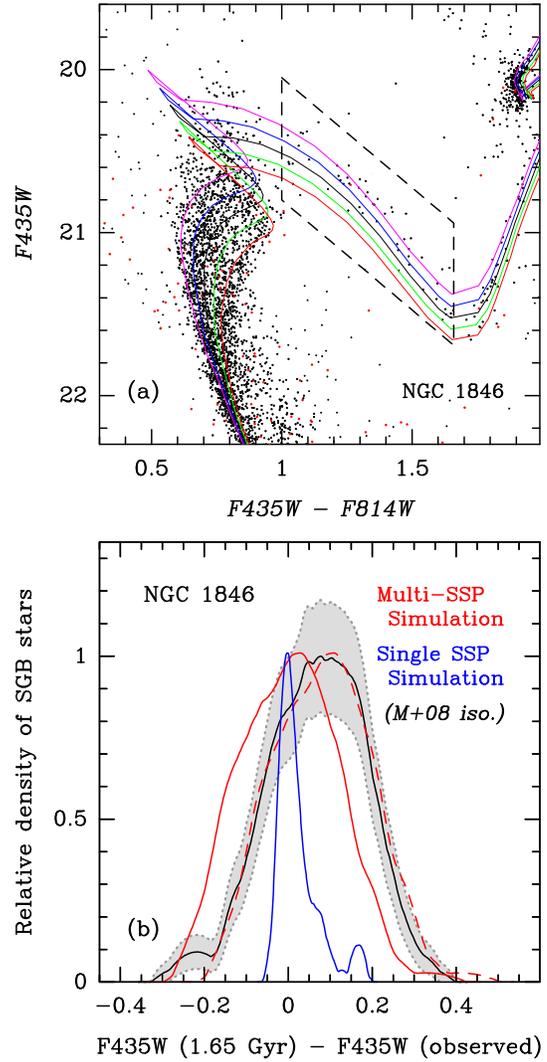}}
\caption{Similar to Figure~\ref{f:SGB1783}, but now for NGC 1846. The SGB
  profile of the ``Single SSP Simulation'' in panel (b) was divided by
  a factor of $\simeq 4.4$ in order to scale its maximum density to
  that of NGC~1846. 
  } 
\label{f:SGB1846}
\end{figure}

We performed a similar analysis of the SGB regions in the massive
eMSTO clusters NGC 1783, NGC 1806, and NGC 1846, using the data from
\citet{goud+09} and G+11a. Since the ages of these three clusters are
very similar to one another, we adopt the M+08 isochrone for an age of
1.65 Gyr to set the magnitude  zeropoint for the cross-SGB profiles of
these clusters. The results are shown in Figs.\
\ref{f:SGB1783}\,--\,\ref{f:SGB1846}. Similar to the case of NGC~1651,
we find that the shapes of the cross-SGB profiles of all three
clusters are consistent with those of their respective multi-SSP
simulations, while they are significantly wider than those of their
single-SSP simulations. And again, the clusters' cross-SGB profiles
are offset from those of their multi-SSP simulations by a shift of
$\simeq$\,0.08 mag.  
Similar to the interpretation of this offset by L+14 for the case of NGC~1651,
BN15 interpreted this offset for the cases of NGC 1806 and NGC 1846 in the sense
that their SGB morphologies are inconsistent with the presence of a significant age
spread\footnote{Note however that the \emph{shapes} of the SGB profiles of NGC
  1806 and NGC 1846 shown by BN15 (see their Figs.\ 3 and 8) seem to be more
  consistent with their SSP simulations that involve a spread of ages than with
  those of a single-age SSP.}.

\begin{figure}
\centerline{\includegraphics[width=7.0cm]{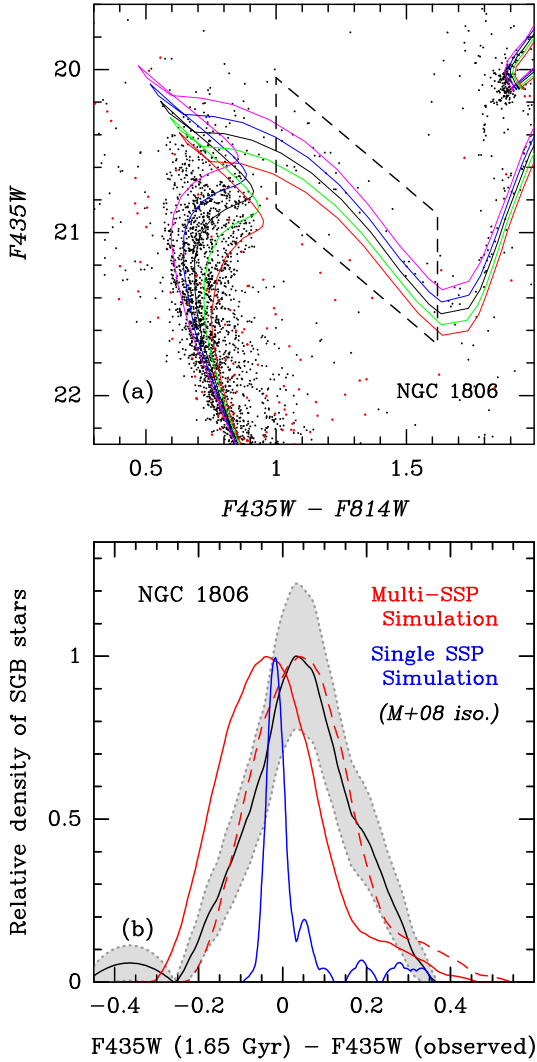}}
\caption{Similar to Figure~\ref{f:SGB1783}, but now for NGC 1806. The SGB
  profile of the ``Single SSP Simulation'' in panel (b) was divided by
  a factor of $\simeq 4.6$ in order to scale its maximum density to
  that of NGC~1806. 
  } 
\label{f:SGB1806}
\end{figure}

\subsubsection{Summary}

Summarizing the results in this Section, it seems that the \emph{shapes} of
cross-SGB profiles of eMSTO clusters are consistent with those of their
respective multi-SSP simulations, while they are inconsistent with those of
their single-SSP simulations. 
Note that in this sense, the cross-SGB profiles of such clusters are entirely
consistent with their cross-MSTO profiles when the latter are interpreted as a
distribution of ages (i.e., the pseudo-age distributions derived by G+11a and
G14).  

The main inconsistency between the cross-SGB profiles of eMSTO clusters and
those of their respective multi-SSP simulations shown above is that of an
magnitude offset of $\approx$\,0.1 mag between them (for the filters
of the \HST\ instruments used here, i.e., F475W of WFC3 and F435W and F555W of
ACS/WFC) in that the clusters' SGB stars are brighter than the simulations
that use their best-fitting M+08 isochrones. This issue is addressed in the
next Section.

\section{Discussion}
\label{s:disc}

\subsection{The Impact of Convective Overshoot}
\label{s:overshoot}

Stars in the MSTO and SGB region of the CMD for ages of $\sim$\,1.5\,--\,2.5
Gyr have masses in the range $1.3 \la (\cM/M_{\odot}) \la 1.6$. 
One parameter in stellar evolution theory that is known to have a significant
impact on the photometric properties of stars in this mass range 
is that of the level of ``overshoot'' from the convective core, i.e., the mean
free path of convective bubbles across the border of the convective region
\citep*[e.g.,][]{mead75,bres+81,bert+85,dema+04}. The level of convective
overshoot $\lamc$ depends on stellar mass in the mass range 
between fully radiative cores ($\cM \la 1.0\; \Modot$) and that of fully
convective cores ($\cM \ga 1.5\; \Modot$), and it is also believed to depend on
metallicity \citep{dema+04,bres+12}. 
The detailed dependencies of $\lamc$ on stellar mass and metallicity are not
well-established, due in part to a relative lack of high-quality data of
massive intermediate-age star clusters for calibration purposes. A consequence
of this uncertainty is that different SSP models implement the dependencies
of $\lamc$ on stellar mass and metallicity in different ways \citep[compare,
e.g.,][]{piet+04,dott+08,mari+08,bres+12}. 

In the following we investigate whether an adjustment of the level of
overshoot can reasonably account for the observed magnitude offset between the
SGB profiles of clusters and those of their SSP simulations. To do so, we run
a set of isochrones based on PARSEC tracks \citep{bres+12} 
using a range of $\lamc$. The model set covers 
$0.30 \leq \lamc \leq 0.70$, where values for $\lamc$ are in the
\citet{bres+81} formalism\footnote{$\lamc$ refers to the maximum 
    level of overshoot in the relevant stellar mass range. The increase of
    overshoot level with stellar mass is assumed to be linear.}, 
for a range of ages and metallicities relevant to the clusters presented in this
paper. 
Note that there are additional ways to adjust the level of overshooting for a 
given isochrone, such as shifting the stellar mass range in which overshooting
takes place or by adopting a diffusive approach for this process 
\citep[e.g.,][]{frey+96,herw00}.

\begin{figure}
\centerline{\includegraphics[width=7.0cm]{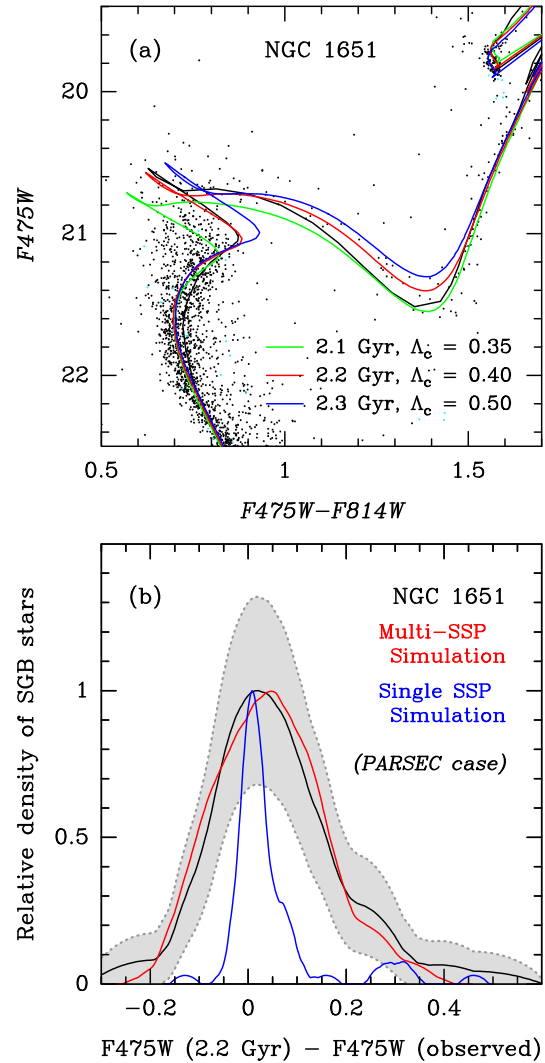}}
\caption{\emph{Panel (a)}: CMD of NGC 1651 showing the upper MS, MSTO, SGB,
  lower RGB, and Red Clump features. The black line represents the
  best-fitting \citet{mari+08} isochrone (cf.\ Fig.\ \ref{f:n1651cmds}a). The
  green, red, and blue lines represent PARSEC isochrones with [M/H] =
  $-$0.40 for which the ages and overshoot values are indicated in the legend. 
  \emph{Panel (b)}: Same as Fig.\ \ref{f:SGB1651}, but now using the SGB
  of the PARSEC isochrone with age = 2.2 Gyr and $\lamc = 0.40$ as the
  zeropoint of the cross-SGB profiles. Note the similarity of the cross-SGB
  profiles of NGC~1651 (black line with 68\% confidence interval in light
  grey) and that of its ``multi-SSP simulation'' (red line), while the
  cross-SGB profile of the ``single-SSP simulation'' (blue line) is much
  narrower. 
  } 
\label{f:overshoot_1651}
\end{figure}

Fig.\ \ref{f:overshoot_1651}a shows the CMD of NGC 1651 along with three 
PARSEC isochrones with levels of overshoot ranging from $\lamc$ = 0.35 to
0.50. For each case, the age 
of the isochrone was adjusted to coincide with the best-fitting isochrone from
the M+08 models (shown in Fig.\ \ref{f:n1651cmds}, and with a black solid line
in Fig.\ \ref{f:overshoot_1651}) near the red end of the turnoff region (at
${\rm F475W} \simeq 21.15$ and $\gI \simeq 0.82$ to be precise). 
The values for $(m-M)_0$ and $A_V$ were chosen to reproduce the positions of
the RGB, the lower MS, and the red clump (RC), in a way equivalent to that
described by G+11a.  
Note that the shape and location of the top of the MSTO of the PARSEC
isochrone with $\lamc = 0.40$ are virtually identical to those of the
best-fitting M+08 isochrone, \emph{while its SGB is clearly brighter} (by up to
$\sim$\,0.12 mag at the red end of the SGB). Apart from this difference at the
SGB and a small colour difference at the bottom of the MSTO, the PARSEC
isochrone with $\lamc = 0.40$ provides a fit to the CMD features of NGC 1651
that is very similar to that of the best-fitting M+08 isochrone. 
Fig.\ \ref{f:overshoot_1651}b illustrates the impact of the use of this PARSEC
isochrone with $\lamc = 0.40$ for creating cross-SGB profiles of NGC 1651 and
its SSP simulations. This plot represents a copy of Fig.\ \ref{f:SGB1651}
except that the zeropoint of the SGB magnitude offsets is now set by the
PARSEC isochrone rather than the best-fitting M+08 isochrone. \emph{Note
that the cross-SGB profile of NGC~1651 is now fully consistent with that of
its multi-SSP simulation \emph{(i.e., without any residual shift)},
  while its single-SSP simulation is still clearly narrower}. 

Given this result, we suggest that the observed magnitude offsets between the
SGB profiles of star clusters and those of their SSP model isochrones or
simulations can be understood entirely by details of the way the dependencies
of $\lamc$ on stellar mass and metallicity are implemented in the SSP model
being used. As such, we argue against the use of the presence of such 
(small) magnitude offsets between cross-SGB profiles to confirm or deny the
presence of a given age distribution in intermediate-age star clusters such as
the ones discussed in L+14, BN15, and the current paper. 
Instead, we suggest that comparisons between SGB properties of such clusters
and their SSP model predictions be focused on the \emph{shapes} of their
cross-SGB profiles, for clusters with statistically sufficient numbers of
stars on the SGB, and that unresolved binaries are properly taken into
account in such analyses.  

\subsection{The width of the Red Clump as a diagnostic for age spreads}
\label{s:RCwidth}

In addition to their study of the shapes of the SGB, BN15 also presented a
comparison between the RCs in NGC 1806 and NGC 1846 and those of M+08 isochrones
of different ages (their Figs.\ 4 and 9), concluding (1) that the observed RC's
showed a smaller spread in colour than expected from clusters having a
significant age spread, and (2) that the isochrones that fit the
youngest part of the eMSTO were also those that showed the best fit to the mean
RC position. 

To put BN15's analysis in context, we call attention to
a few points that are relevant to the interpretation of the RC's position
and width. First, the (mean) position of the RC depends on a number of
factors in addition to age, distance, and reddening. 
One such factor is the mixing-length parameter $\alpha_{\rm MLT}$, which defines
the efficiency of energy transport in the external layers of the star and hence
its effective temperature $T_{\rm eff}$ for a given luminosity. In most
isochrone sets, including the M+08 ones, $\alpha_{\rm MLT}$ is calibrated in the
solar model, and then applied to all stars including the red giants. Recent 3-D
model atmospheres from \citet{tram+14} indicate however that while the
approximation of a nearly constant $\alpha_{\rm MLT}$ might be good for stars
along the upper RGB (and probably also for the RC), it also 
indicates significant variations between the $\alpha_{\rm MLT}$ values of red
giants and those of dwarfs such as those populating the MS of Magellanic Cloud
clusters. The variations are such (a few tenths in $\alpha_{\rm MLT}$) 
that they could change the relative colours of MS and RC stars by 
a few hundredths of a mag. 
Another uncertainty in isochrone colours is due to the 
adopted $T_{\rm eff}$\,--\,colour relations, which are usually derived from 1-D
static model atmospheres \citep[e.g.,][]{caskur03}. Therefore, one should be
cautious in interpreting issues with regard to ages of star clusters based on 
differences between the MS and RC positions in the CMD, relative to isochrones,
at levels of 0.01\,--\,0.03 mag. 

\begin{figure*}
\centerline{\includegraphics[width=14.0cm]{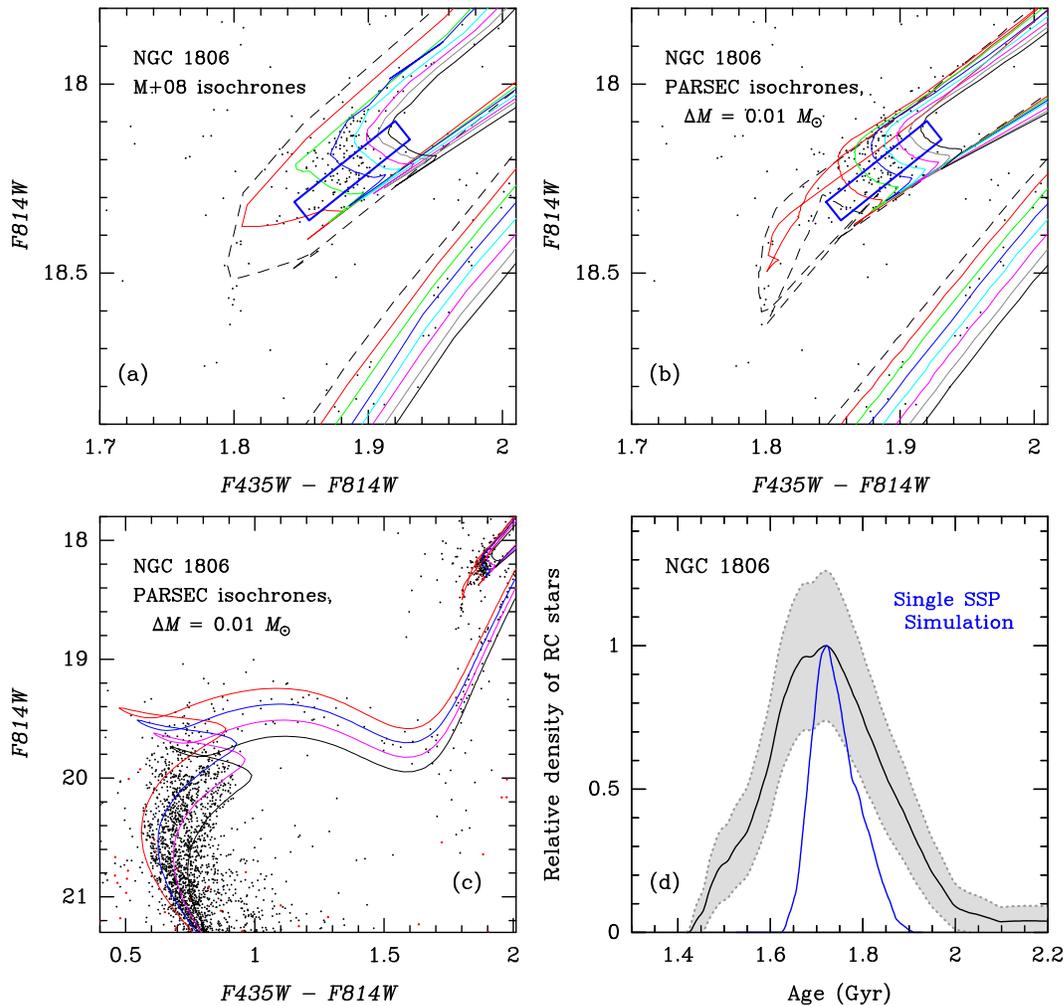}}
\caption{ 
  \emph{Panel (a)}: The CMD of NGC~1806 (cf.\ Fig.\ \ref{f:SGB1806}a),
  zoomed in on the RC region. 
  The lines represent \citet{mari+08} isochrones for log (age/yr) = 9.10
  to 9.24 (left to right) with a spacing of 0.02. 
  \emph{Panel (b)}: Same as panel (a), but the lines now represent new
  PARSEC isochrones with $Z = 0.008$ which were derived from stellar tracks
  spaced by $\Delta \cM/\Modot = 0.01$ (see Sect.\ \ref{s:RCwidth}). Isochrones
  are for log (age/yr) = 9.18 to 9.32 (left to right) with a spacing of 0.02. 
  The thick blue rectangle in panels (a) and (b) is used to derive the age
  distribution of RC stars as described in Sect.\ \ref{s:RCwidth}. 
  \emph{Panel (c)}: CMD of NGC~1806 showing the region from below the MSTO to
  beyond the RC along with new PARSEC isochrones for $Z = 0.008$ and 
  ages of 1.58, 1.74, 1.91, and 2.09 Gyr (left to right). 
  \emph{Panel (d)}: The black solid line shows the cross-RC distribution 
  of NGC 1806 expressed in age units (see Sect.\ \ref{s:RCwidth}), while the
  light grey region indicates its 68\% confidence interval. For comparison, the
  solid blue line represents the cross-RC distribution of a single-SSP
  simulation of NGC 1806, divided by a factor $\simeq 1.9$ to yield
  the same maximum star density as that of NGC 1806 itself. 
  } 
\label{f:RCplot}
\end{figure*}

As to the width of the observed RC, we agree with BN15 that it is compact in
both NGC 1806 and NGC 1846, with its main component spanning just $\approx
0.5$~mag  in F435W\,--\,F814W colour. However, both NGC 1806 and NGC 1846
also exhibit a number of secondary RC stars 		
(hereafter SRC; \citealt{gira+09}; G+14). These are 
He-burning stars that were massive enough to avoid electron degeneracy
settling in their cores after leaving the main sequence. In both clusters, the
SRC stars define an almost vertical feature to the bottom left of the main RC,
centered at ${\rm F435W} - {\rm F814W} \simeq 1.80$ mag, and with F814W magnitudes between
18.4 and 18.7 (see Fig.\ \ref{f:RCplot}); they are seen also in BN15's plots
and are somewhat better delineated in NGC 1806  
than in NGC 1846. These SRC sequences are,
by themselves, a clear indication that \emph{the RC in these clusters is not as
  homogeneous and compact} as claimed by BN15. In the age-spread interpretation
of eMSTOs, the presence of a SRC is simply reflecting the spread in ages (i.e.,
initial stellar masses) of stars leaving the MSTO, and indeed the fraction
  of RC stars in the SRC among eMSTO clusters correlates strongly with the
  fraction of MSTO stars at the youngest (brightest) end of the MSTO 
(G+14). In the rotating scenario, instead, SRC stars are the putative progeny of
fast MS rotators, since rapidly rotating stars have larger core masses at
  the end of the MS era than do nonrotating stars
  \citep[e.g.,][]{maemey00,egge+10}. However, this does not seem to explain the
aforementioned observed correlation with the brighter part of the eMSTO
(see G+14 for more details).   

That said, it is interesting to note that such SRCs in 
NGC 1806 and NGC 1846 
apparently contrast with comparatively narrow ``main'' RCs. As
noticed by BN15, M+08 isochrones of ages between 1.3 and 1.8 Gyr
span a significant range of RC colours 
($\Delta({\rm F435W} - {\rm F814W})\simeq0.1$~mag), which seems wider than the 
RC colour spreads in their Figs.\ 5 and 10. 
There are however two additional points to be taken in consideration in this
context: firstly, BN15 inferred the age distribution from the stars in the RC by
assigning ages based on the ``closest isochrone'' rather than using a smooth
function. This method introduces an intrinsic binning in the age distribution,
with a bin size depending on the age spacing of the isochrones. Secondly, the
M+08 isochrones were created by the interpolation of evolutionary tracks  
(originally taken from \citealt{gira+00}) which, for the mass interval into
consideration ($\cM \simeq 1.5-1.8 \; M_{\odot}$), were computed for initial masses
spaced by $\Delta \cM = 0.05\; M_{\odot}$. 
The issue with such isochrones in the context of the interpretation of the RC
morphologies of these star clusters is that they cannot reveal any CMD structure
caused by variations in stellar evolutionary behaviour that occur over 
a mass interval smaller than $0.05 \; \Modot$. If any such variations occur, the
isochrones' behaviour as a function of age will reflect the \emph{smooth
  interpolation} between the confining tracks, rather than the real expected
behaviour. This is what we believe is happening with the M+08
isochrones.  

As recently noticed by \citet[][see their appendix]{gira+13b}, the onset of
electron degeneracy in stellar cores at $\cM \la 1.7\; M_\odot$ (with the exact
limiting value depending on overshooting efficiency) produces a marked
discontinuity in the evolutionary behaviour of stars, occurring over a mass
scale smaller than $0.01\; M_{\odot}$. This significantly affects the expected
shape of the RC in more detailed isochrone models, where stars are in principle 
expected to be either on the main RC or in the SRC, but rarely in between these
features (modulo unresolved binary stars). As a consequence, the
RC is expected to be more compact than that displayed in M+08 isochrones. 
The expectation from PARSEC isochrone models computed from tracks with a fine grid
of stellar masses -- namely those derived from a $Z=0.008$ grid with
the same level of detail and mass resolution as in \citet[][i.e.,
$\Delta \cM/\Modot = 0.01$]{gira+13b} -- is shown in   
Fig.\ \ref{f:RCplot}a, and compared with M+08 isochrones 
(Fig.\ \ref{f:RCplot}b) using the same values for $(m-M)_0$ and
$A_V$. In both cases, we superimpose the stars observed within the
core radius of NGC 1806.  
It can be seen that, with the new PARSEC models, the extent of the
observed ``main'' RC appears to be compatible with an age range of 
$\Delta {\rm log\,(age)} \simeq 0.10$ (or $\Delta {\rm (age)} \simeq 500$ Myr
at an average age of 1.7 Gyr), whereas it indicates a smaller age
range in the case of the M+08 isochrones ($\Delta {\rm log\,(age)} \simeq 
0.07$ or $\Delta {\rm (age)} \simeq 200$ Myr).  

The ability of the new PARSEC isochrone models that were computed from a grid
of initial stellar masses spaced by $\Delta \cM = 0.01\; M_\odot$ to fit the
full CMD of NGC~1806 is illustrated in Fig.\ \ref{f:RCplot}c. We used
$(m-M)_0$ = 18.53 mag and $A_V$ = 0.035 mag for these isochrones. Note
the good fit to the MSTO, SGB, RGB, and RC regions. 

Finally, we determine the age distribution from the RC of NGC~1806 by 
means of the distribution of stars in a rectangle in the RC region of the F814W
vs. F435W\,--\,F814W CMD. 
The short and long axes of the rectangle are approximately parallel and
perpendicular to the isochrones, respectively (see blue rectangle in Fig.\
\ref{f:RCplot}b).  
The magnitudes and colours of stars in the rectangle are then transformed into
the coordinate frame defined by its two axes. The star positions in the 
coordinate perpendicular to the isochrones are then translated to age by
repeating the same procedure for the new PARSEC isochrone tables described
above, for an age range that fully covers the observed extent of the RC region
of NGC~1806 (using the same values of $Z$, $(m\!-\!M)_0$ and $A_V$), and
performing a polynomial least-squares fit between age and the coordinate
perpendicular to the isochrones\footnote{We excluded the SRC area
  of the CMD from this rectangle to avoid the highly non-linear (and
  multi-valued) functional relation between age and the coordinate along the
  long axis of such a rectangle in the SRC region. This causes an 
  underestimate of the number of stars at the youngest ages shown in Fig.\
  \ref{f:RCplot}d.}. The resulting age distribution of NGC~1806 is shown in
Fig.\ \ref{f:RCplot}d (black line), using a non-parametric Epanechnikov density
kernel as before. For comparison, Fig.\ \ref{f:RCplot}d also shows the age
distribution that would be expected for a single SSP, derived from a set of
Monte Carlo simulations (as described in Sect.\ \ref{s:meth}) using the new
PARSEC isochrone for an age of 1.74 Gyr.  
Note that while the age distribution is shifted to slightly older ages
relative to those indicated by the cross-MSTO profiles when the latter are
compared with M+08 isochrones (cf.\ Fig.\ \ref{f:weights}c), the cross-RC
profile shape of NGC 1806 is similar to that of its cross-MSTO one, and it is
significantly wider than that of a single-SSP simulation.  

Although these new RC models will be described in more detail in a forthcoming 
paper, these results already demonstrate that the expected detailed RC
morphology of intermediate-age clusters that host a range of stellar ages is
model --and mass-resolution-- dependent. We therefore argue that the
conclusions by BN15 on the nature of the RCs of NGC~1806 and NGC~1846 
were largely due to their choice of isochrones and analysis methods. In
contrast, our results with newly updated isochrone models strongly suggest that
the RC morphologies of these clusters are actually \emph{described very well by a
distribution of stellar ages that is consistent with that indicated by their MSTOs}.

\subsection{Implications regarding the nature of eMSTOs}
\label{s:impl}

Our result that the widths and shapes of cross-SGB profiles of eMSTO clusters
are consistent with those of their respective multi-SSP simulations, and
significantly wider than those of their single-SSP simulations, has a
number of interesting implications related to the nature of eMSTOs. 

It is obvious that this result is consistent with the prediction of the
``extended star formation'' scenario; we therefore focus on a comparison with
predictions of the ``stellar rotation'' scenario. It is well known that the
centrifugal force in rotating stars with masses in the approximate range of
1.2\,--\,1.7 $\Modot$ decreases their effective gravity which in turn lowers
their effective temperature 
\citep[e.g.,][]{meymae97}. This prompted \citet{basdem09} to suggest that a
range of stellar rotation velocities may cause the eMSTO feature. However,
\citet{gira+11} computed an isochrone for a typical age of eMSTO clusters and
a typical stellar rotation velocity and found that the prolonged lifetime of
rotating stars approximately cancels the effects of the lower gravity
in terms of the MSTO colour, which was found to be virtually
equal to that of the non-rotating isochrone. The main impact of stellar
rotation was found to be a \emph{lengthening} of the MSTO hook in conjunction
with a brightening of the SGB (see Fig.\ 3 in \citealt{gira+11}), similar to
the effects of an increased level of convective overshoot shown in
Sect.\ \ref{s:overshoot}. 

More recently, \citet{yang+13} computed isochrones that included the effects
of stellar rotation, using the Yale Rotating Evolution Code
\citep{pins+89}. They found that rotation can actually cause eMSTOs if the
efficiency of rotational mixing is sufficiently small\footnote{Note however
  that G+14 pointed out that the width of the MSTO of the most massive eMSTO
  clusters is significantly larger than the predictions of \citet{yang+13},
  even for their smallest value of rotational mixing efficiency.}. However,
the luminosities of the onset of the SGB of the rotating isochrones was found to
be consistent with those of the corresponding non-rotating isochrones. 
This was 
also part of the arguments of L+14, who used the recent tracks of
\citet{geor+14} and showed that the track for a rapidly rotating star is at a
cooler temperature than that of the non-rotating track at the MSTO, while the
rotating track merges with the non-rotating track right at the onset of the
SGB. 
However, we remind the reader that due to the longer lifetime of rotating
stars, one should not use stellar tracks to interpret the morphologies of the
MSTO or SGB in CMDs. One should compute isochrones instead. This will be
addressed in a forthcoming paper.

In summary, a spread of rotational velocities in a single-age stellar
population would cause the MSTO and SGB features to show the following
properties according to the predictions of the recent modeling efforts: \\ [-3.8ex]
\begin{itemize}
\item According to the findings of \citet{gira+11}, the MSTO would
  not show any significant spread in colour, only a spread of the
  \emph{length} of the MSTO hook. This is inconsistent with the MSTO
  morphologies of eMSTO clusters. Conversely, the SGB \emph{would} show a
  significant spread of luminosities (similar to what we find in eMSTO
  clusters).  
\item According to \citet{yang+13} and L+14, the MSTO would show some
  significant spread in colour (similar to the morphology of eMSTOs) if the
  efficiency of rotational mixing is small, but the 
  luminosity of the SGB would \emph{not} be affected in a significant way. The
  latter prediction is inconsistent with our cross-SGB profiles of eMSTO
  clusters.  \\ [-3.8ex]
\end{itemize}

Given the above, our result that the cross-SGB profiles of eMSTO clusters are
significantly wider than that of their single-SSP simulations (and consistent
with the widths of MSTOs of eMSTO clusters) seems to indicate 
that \emph{their MSTO and SGB morphologies are better described by a
  distribution of stellar ages than by the combination of a single age
  and a range of stellar rotation velocities}. We recognize that the
extent to which rapid stellar rotation affects the temperature and
luminosity of SGB stars is still in some level of flux. For example,
the grid of models by \citet{geor+14} has so far only been completed
for stars with $\cM \geq 1.7 \; \Modot$, while SGB stars in the eMSTO
clusters discussed in this paper are less massive (by 0.1\,--\,0.3
$\Modot$)\footnote{We note that the study of \citet{yang+13} 
  showed that the impact of stellar rotation to isochrone shapes changes
  significantly in this mass range.}. As such, our conclusions refer to the
situation at the time this paper is written.   

\section{Summary and Conclusions}
\label{s:conc}

In the context of the question of the nature of eMSTOs in massive
intermediate-age star clusters in the LMC, we investigated the recent claims of 
\citet[][L+14]{li+14} and \citet[][BN15]{basnie15} regarding 
the nature of such clusters' SGB morphologies, which are expected to be
affected more by age spreads than by spreads in stellar rotation velocity. Their 
analysis led them to argue that the SGB morphologies of three clusters featuring
eMSTOs (NGC~1651, NGC~1806, and NGC~1846) are inconsistent with extended star
formation histories within those clusters. 
We performed an independent study of the SGB morphologies of five
intermediate-age star clusters in the LMC, including the three clusters studied 
by L+14 and BN15 (NGC~1651, NGC~1806, and NGC~1846). Comparisons between the
SGB morphologies of cluster stars and those of SSP model predictions are done
using Monte-Carlo simulations of single-age SSPs as well as ``composite SSPs''
whose age distributions are taken from the ``pseudo-age distributions'' of the
clusters which were derived from their MSTO morphologies by G+11a and G+14. The
SSP simulations include a self-consistent treatment of unresolved
binaries. Other methodological differences with the L+14 and BN15
studies are pointed out where relevant. Our main conclusions are the
following. \\ [-3.8ex]   
\begin{enumerate}
\item
In contrast with L+14 and BN15, we find that the shapes of the cross-SGB
profiles of all clusters in our sample are consistent with their
cross-MSTO profiles (to within 1$\sigma$) when the latter are interpreted as
distributions in age. Conversely, cross-SGB profiles of simulated single-age
stellar populations (convolved with photometric uncertainties) are found to be 
significantly narrower than those of the clusters. 
\item 
We argue that the magnitude offset between the cross-SGB distributions of
the clusters in our sample and that of their best-fitting isochrones from the
\citet{mari+08} family of SSP models can be understood once the variation of the
treatment of convective overshoot among SSP isochrone models is taken into
account.   
\item 
We investigate the claims made by BN15 who stated that the red clumps of
NGC~1806 and NGC~1846 showed a smaller spread in colour than expected from
clusters having a significant age spread, and that the M+08 isochrones that fit
the  youngest part of the eMSTO were those that showed the best fit to the mean
RC position. We find that these results are dependent on the models and methods
used. Using isochrones newly created from a grid of PARSEC tracks that 
features a 5-fold higher resolution in stellar mass than that used by the M+08
(and most other sets of) isochrones, we find that ages indicated by the MSTO of
NGC~1806 are actually \emph{consistent} with those indicated by the
RC. Furthermore, we find that the RC morphology of NGC~1806 is consistent with the age
distribution indicated by the MSTO, and significantly more extended
than that expected from a single-age SSP. 
\item
We compare the observed MSTO and SGB morphologies of eMSTO clusters
with those of (current) predictions of models that assume (a) a distribution of
stellar ages versus (b) a range of stellar rotation velocities at a single age. This
comparison indicates that a distribution of stellar ages is currently the only option 
that can explain both the observed MSTO and SGB morphologies. \\ [-3.5ex]
\end{enumerate}
Our overall conclusion is that in spite of recent arguments by 
L+14 and BN15, the SGB and RC morphologies of star clusters featuring eMSTOs are
consistent with the scenario in which eMSTOs are (mainly) due to a distribution
of stellar ages.  

\paragraph*{Acknowledgments.} \ \\ 
This work is based on observations obtained with the NASA/ESA \emph{Hubble
    Space Telescope}, obtained at the Space Telescope Science
  Institute, which is operated by the Association of Universities for
  Research in Astronomy, Inc., under NASA contract NAS5-26555. 
Partial support for this project was provided by NASA through grant HST-GO-12908 
from the Space Telescope Science Institute. 
We are grateful for a very thoughtful review by the anonymous referee,
  which improved this paper. 
P.M. and L.G.\ acknowledge support from the University of Padova (\emph{Progetto
  di Ateneo 2012}, ID: CPDA125588/12), and from the ERC Consolidator Grant funding 
scheme (\emph{project STARKEY}, G. A.\ n.\  615604).   
T.H.P.\ acknowledges support through FONDECYT Regular Project Grant
No.\ 1221005 and BASAL Center for Astrophysics and Associated
Technologies (PFB-06). 
We acknowledge relevant discussions with Vera Kozhurina-Platais. 
We made significant use of the SAO/NASA Astrophysics Data System during this
project, and we acknowledge use of the R Language for Statistical
Computing, see http://www.R-project.org.

\end{document}